# Frictional Duality Observed during Nanoparticle Sliding

Dirk Dietzel[1,2], Claudia Ritter[3], Tristan Mönninghoff[1], Harald Fuchs[1,2], André Schirmeisen[1]*, Udo D. Schwarz[3]

[1] Physikalisches Institut and Center for Nanotechnology (CeNTech), Westfälische Wilhelms-Universität Münster, Münster, Germany
[2] INT, Forschungszentrum Karlsruhe (FZK), Karlsruhe, Germany
[3] Department of Mechanical Engineering and Center for Research on Interface Structures and Phenomena (CRISP), Yale University, New Haven, CT, USA

**One of the most fundamental questions in tribology concerns the area dependence of friction at the nanoscale. Here, experiments are presented where the frictional resistance of nanoparticles is measured by pushing them with the tip of an atomic force microscope. We find two coexisting frictional states: While some particles show finite friction increasing linearly with the interface areas of up to $310,000$ nm$^2$, other particles assume a state of frictionless sliding. The results further suggest a link between the degree of surface contamination and the occurrence of this duality.**

*[E-mail:] schirmeisen@uni-muenster.de

In recent years, considerable efforts have been directed towards the clarification of the atomic origins of friction [1], largely spurred by the ongoing miniaturization of moving components in technological devices and the advent of new techniques allowing nanometer-scale tailoring of tribological surface coatings [2]. One of the most debated subjects in this context is how the frictional force $F_f$ experienced at a finite, atomically flat interface of nanoscopic dimensions scales with the actual contact area $A$. Macroscopically, Amontons' well-known law $F_f = \mu F_l$ applies, where $F_l$ represents the external loading force and $\mu$ the friction coefficient. Since $\mu$ is constant for a given material combination, friction is independent of the dimensions of the interface. If, however, we perform the transition from the *apparent* macroscopic contact area to the *true dimensions* of the nanometer-sized single asperity contacts that actually support the macroscopic sliders, this behaviour changes.

So far, only two studies using realistically sized isolated nanocontacts of some 100,000 nm$^2$ have been published indicating linear dependencies between friction and contact area [3, 4]. Their interpretation, however, is hampered because they have not been performed under well-defined vacuum conditions. In contrast, other reports indicate that virtually frictionless sliding may exist under certain conditions for model contacts of only a few nm$^2$ in size [5, 6, 7, 8]. Indeed, for sliding contacts featuring clean, atomically flat, but incommensurate interfaces, theory predicts 'superlubric' sliding, as the structural mismatch induces a decrease of the barriers between local minima of the interaction potential with increasing contact size that ultimately leads to vanishing friction [9, 10, 11, 12]. This mechanism, which has also been denoted as 'structural lubricity' [9], is widely held responsible for the excellent lubrication properties of solid lubricants such as graphite [7] or molybdenum disulfide [13]. The prospect of establishing an analogue regime with sliders used in micromachines motivates our analysis of the frictional properties of *extended* nanocontacts with varying size under controlled conditions.

The present lack of knowledge in this regard stems from the fact that established experimental procedures are severely limited due to a size gap between the small contact areas of scanning probe microscopes (few nm$^2$) [5, 6, 7, 8] and the contact areas offered by the surface force apparatus (some ten thousands of $\mu$m$^2$) [14, 15]. To overcome this gap, we performed experiments where the frictional resistance of nanoparticles is measured while they are pushed by the tip of an AFM. Nanoparticle manipulation has been used to investigate different aspects of friction (e.g., rolling versus sliding [16]), but so far rarely for quantitative, statistically significant size-dependent studies [3, 4]. Fig. 1(a) illustrates how particle translation has been realized in the present work; the applied data acquisition and analysis procedures have been developed specifically for this investigation (cf. [17]). Basically, quantitative values for the particle's frictional resistance are extracted from individual line traces of the friction signal acquired during manipulation (Fig. 2).

As sample, we chose antimony grown on highly oriented pyrolytic graphite (HOPG) (see Fig. 1(b)). This material combination is well suited for manipulation experiments [4], and its growth and structure have been characterized before [18]. If not stated otherwise, HOPG substrates were cleaved under ambient conditions, immediately introduced into the vacuum chamber (p $< 5\times10^{-10}$ mbar) and subsequently heated in-situ to $150°$ C for 1 h in order to ensure clean surface conditions. Antimony was evaporated from the solid phase at $370°$ C for 20 minutes. A low evaporation rate was chosen, which resulted in mostly round or only modestly ramified islands of 50-750 nm in diameter and up to 80 nm in height. Cross-sectional TEM studies revealed flat particle surfaces at the HOPG-Sb interface with no indications of asperity formation. All four independent data sets presented have been obtained at room temperature. For data sets #1 and #2, experiments were carried out using Omicron Nanotechnology's standard ultrahigh vacuum (UHV) AFM system, while set #3 was recorded using their VT-series UHV-AFM. In both cases, the sample was transferred from the preparation stage to the AFM without breaking the vacuum. Complementary experiments were executed under ambient conditions using a Veeco Multimode AFM (set #4).

Data sets #1 and #2 represent an original experiment (circular markers in Figs. 3(a) and (b)) and a control experiment (square markers) that have been performed with two different cantilevers and different, but identically prepared samples. The results of 31 dislocation events using particles featuring contact areas between $22,000$ nm$^2$ and $90,000$ nm$^2$ are presented in Fig. 3(a). Contact areas are defined as the cross-sectional areas determined from the AFM topography images [4]. These events can be categorized in two distinct regimes: While the majority featured substantial frictional resistance (regime 1; black symbols), about one quarter of the events showed almost no detectable friction (regime 2; red symbols), causing an apparent 'frictional duality'.

To clarify the friction-size relation for the events with substantial frictional resistance, we extended the experiments by including particles up to $A = 310,000$ nm$^2$. The results shown in Fig. 3(b) suggest a linear dependence and a constant shear stress $\tau = F_l/A = (1.04 \pm 0.06)$ MPa. Since the normal force experienced by the particles is due to adhesion, which scales linearly with area, an area-independent friction coefficient follows, reinforcing Amontons' law also at the nanoscale. Thereby, $\tau$ is almost identical to values found for Cd arachidate islands ($1$ MPa) [19] and MoO$_3$ nanocrystals ($1.1$ MPa) [3] moved at ambient pressure, but roughly one order of magnitude higher than for C$_{60}$ islands displaced on NaCl in UHV [20]. Also note that vanishing friction events were only found for islands smaller than $90,000$ nm$^2$, which may indicate that strain relaxations or even grain boundaries of larger islands could play a role.

To discuss the observed coexistence of two frictional states, let us consider four different scenarios: (i) Particles showing no apparent friction are picked up by the tip during translation. (ii) Particles showing no apparent friction are stuck on a graphite flake, which slides superlubric [7]. (iii) Particles are crystalline and exhibit well-ordered, crystalline interfaces. Depending on the particle lattice's orientation relative to the substrate, finite friction (commensurate) or vanishing friction (incommensurate) will be observed [5, 7, 12]. (iv) While clean interfaces may exhibit superlubric behavior due to structural mismatch (particles may be crystalline or amorphous), even small amounts of mobile molecules (such as hydrocarbon or water molecules) trapped between the sliding surfaces can cause a breakdown of the superlubric behavior [9, 21]. Often referred to as 'dirt particles', these molecules are able to move to positions where they simultaneously match the geometry of both top and bottom surfaces, thus augmenting the height of the bottom surface in a way that matches the (atomic-scale) undulations of the top surface [9]. As a consequence, an area independent friction coefficient is obtained.

From these scenarios, the pick-up hypothesis (i) can be discarded since we could image frictionless displaced particles after translation (see Fig. 4). Also, it seems energetically not feasible for a tip to lift off a plate-like particle from the surface that is adhesively bound to the surface by an area much larger than any of its side faces. Next, we consider the case of a graphite flake stuck underneath the particle (ii) being unlikely since the images recorded after particle manipulation show no sign of missing graphite flakes. We have also observed that two initially superlubric manipulation events (see below) turned into 'regular' events featuring finite friction after a short travel distance, which would be difficult to understand in this picture.

Next, we note that the compact shape of most of the particles used for sets #1 and #2 suggests them being amorphous [18], while the scenario of commensurate crystalline interfaces (iii) requires the presence of a commensurate crystalline interface. And if indeed some of the particles were crystalline, the atomic lattices of Sb and HOPG do not match, i.e., the interfaces are incommensurate and superlubric behavior should prevail (even though the existence of Moire-type potential minima for certain relative lattice orientations may be conceivable). In any case, if we assume the model of commensurate surfaces being correct, it would still be surprising that the superlubric state only occurs for one quarter of the investigated islands, while most particles still exhibit 'Amontons-like' sliding. Please note in this context that multiple translation of the same particle (up to 22 subsequent manipulations have been performed with the same particle) did not lead to a statistically relevant change of the observed frictional force (cf. Fig. 8 in [17]), even though small particle rotations would already result in vanishing friction. These arguments make lattice orientation-dependent friction as assumed in scenario (iii) appear rather unlikely.

The 'dirt particle' scenario (iv) finally assumes that adsorbed but mobile molecules are trapped between the sliding surfaces. Even under 'clean' UHV conditions, a fair number of such mobile adsorbates can accumulate on HOPG surfaces over extended periods of time (sets #1 and #2 were collected over several weeks). Further, the HOPG samples for these data sets have been cleaved in ambient conditions, and the subsequent in-situ cleaning by heating might have been insufficient. Therefore, we improved the experimental procedure by cleaving the HOPG crystal inside the UHV chamber and conducting all experiments within two days while focusing on smaller islands ($< 40,000$ nm$^2$). As a consequence, the ratio of particles showing no friction was greatly increased to over one half of the manipulated particles in the corresponding set #3 (triangles in Fig. 3(a)).

In addition, the opposite case of very contaminated surfaces was investigated in a fourth experimental series performed under ambient conditions (Fig. 3(c)). Friction now increased by a factor of 40, and the vast majority of islands now exhibited 'Amontons-like' sliding as found earlier [4]. A parallel analysis of the particles by cross-sectional transmission electron microscopy revealed that the exposure to air converts the particle's surfaces to amorphous antimony oxide, disregarding whether a particle has originally been amorphous or crystalline. While this structural transition might at least partially explain the huge change in frictional resistance, it also implies that all particles should slide frictionless due to the structural mismatch, which is clearly not the case. Nevertheless, we found two events, recorded shortly (hours) after exposing the sample to air that exhibited vanishing friction. Interestingly, both particles involved into these events translated only less than 100 nm superlubric, then converting into 'regular' particles exhibiting substantial friction upon subsequent manipulation.

In conclusion, we find that among the four discussed scenarios, only scenario (iv) remains without apparent contradictions. Further, we could successfully link the ratio of superlubric to non-superlubric events to sample cleanliness and observed the transition from superlubric to 'regular' sliding for particles translated in air. However, additional experiments would be necessary to actually proof or disproof the 'interfacial mobile molecule' hypothesis of scenario (iv); possible alternative mechanisms could include strain relaxations, interparticle grain boundaries, interfacial defects, or effects involving mobile molecules at the particle perimeter. Besides, it seems surprising that the experienced finite friction is very reproducible in all experiments. Intuitively, one would expect a certain variation with degree of contamination especially for the low-contamination experiment (data set #3), even though simulations see very little influence of the level of contamination on friction for a coverage between one quarter and a full monolayer [21]. Also, these predictions might need to be modified for the present case of a pure adhesive load. In any case, the ambient manipulation experiments prove that the superlubric state can, if only rarely and for short distances, be observed even under strongly contaminated conditions, which gives hope for future technical applications of frictionless sliding.

The authors thank H. Hölscher, M. Heyde and M. Müser for discussions. Work at Münster was supported by the DFG (grant No. SCHI 619/8-1) and the FANAS Initiative of the European Science Foundation (CRP 'Nanoparma'), work at Yale by the Petroleum Research Fund of the American Chemical Society (grant No. PRF 42259-AC5) and the National Science Foundation (grant No. MRSEC DMR 0520495). C. R. acknowledges the receipt of a personal stipend by the DFG.

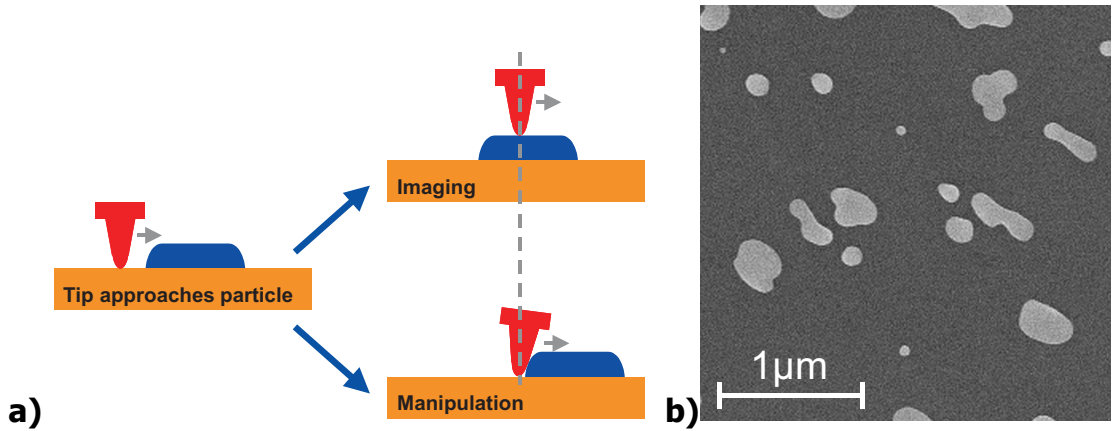

Figure 1: (a) Scheme of particle manipulation experiments. *Imaging*: Below a certain normal threshold force, the cantilever traces the topography accurately without moving the nanoparticle. *Manipulation*: At loads larger than manipulation threshold, the tip pushes the particle out of its way. In this case, an additional lateral force manifesting as enhanced cantilever torsion can be observed, which corresponds to the particle's frictional resistance. (b) Scanning electron micrograph of a sample (Sb on HOPG) used for the experiments.

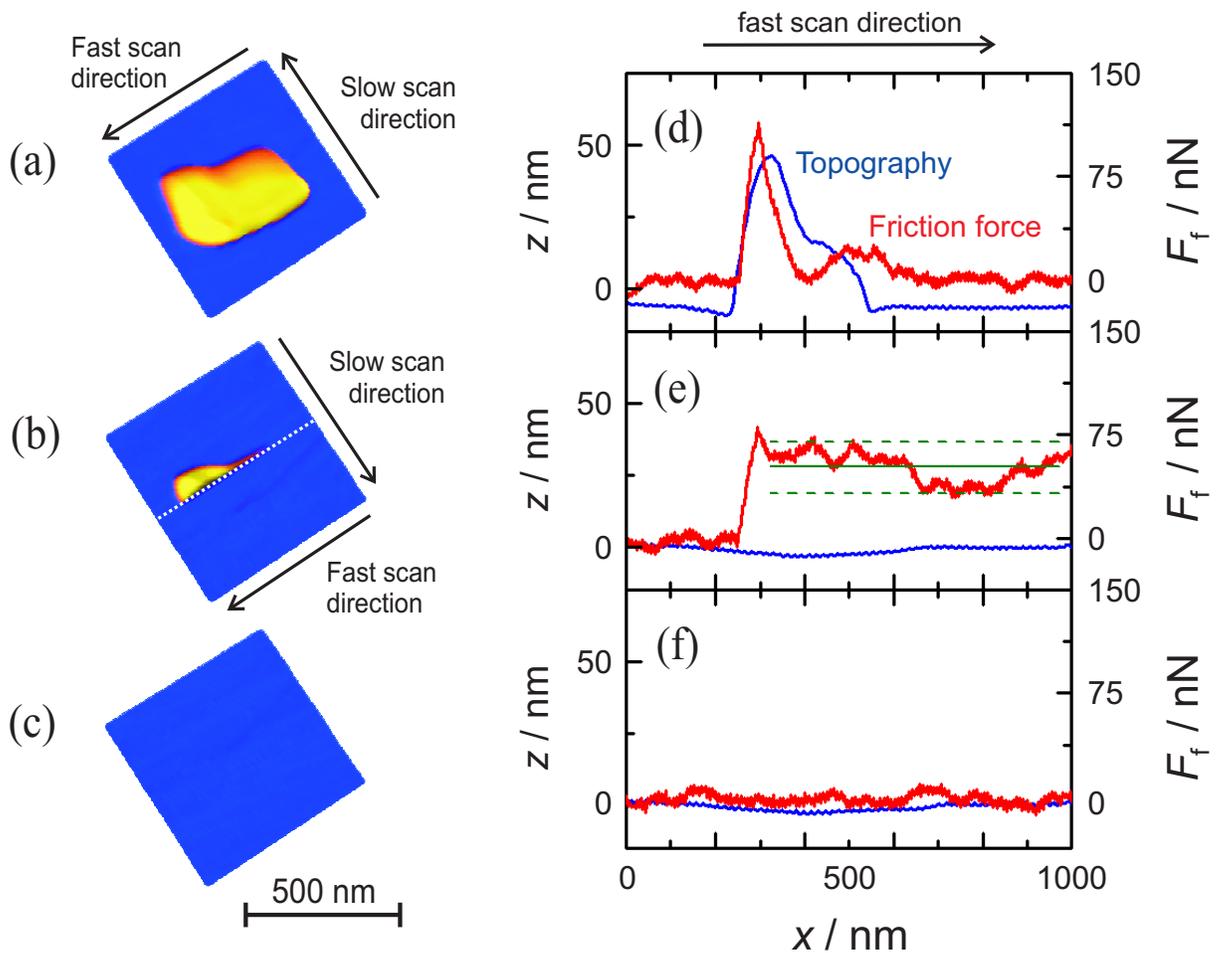

Figure 2: Illustration of the manipulation procedure. (a) A nanoparticle is imaged for contact area determination at low external loading force. (b) The load is increased slightly above the manipulation threshold. Since the force necessary to initiate particle motion for the first time is typically higher than in subsequent manipulation events, the particle is imaged for several line scans before it is pushed out of the field of view (along white dotted scan line), thus showing a 'cut' particle. (c) Subsequent imaging with low loads confirms successful particle translation. (d) Topography (blue, left axis) and lateral force (red, right axis) of the last scan line before translation. The lateral force signal is mainly topography-induced, as the cantilever twists at the particle's edges. (e) Scan line during displacement. The topography now reflects the flat graphite surface, while the average frictional resistance of the particle ($\approx 60$ nN, solid green line) can be determined from the lateral force signal. Fluctuations in the force signals did not show systematic trends and are treated as statistical errors (dashed green lines). (f) First scan line following the manipulation event proving that the particle has been removed.

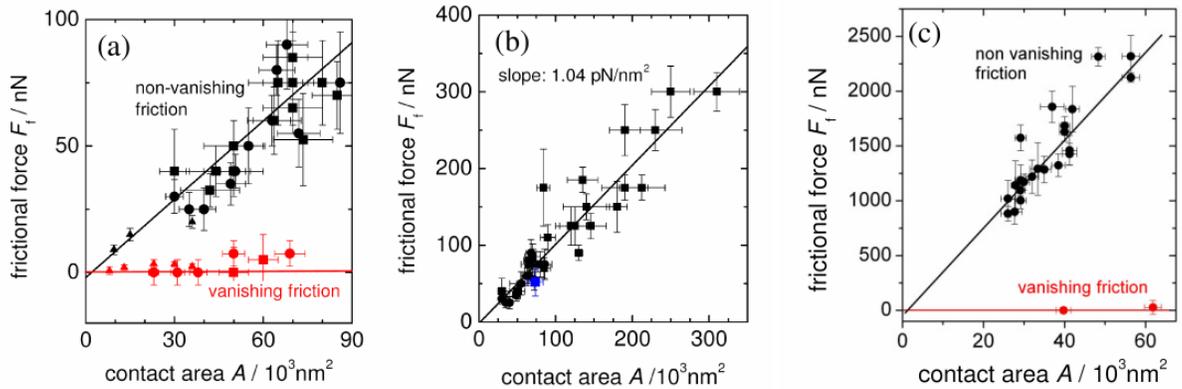

Figure 3: (a) Data obtained under UHV conditions, uncovering two distinct frictional regimes for particle/sample contact areas up to $90{,}000~\text{nm}^2$. Regime 1 (black symbols) comprises particles with substantial friction whereas particles that exhibit virtually no measurable friction (red symbols) are assigned to regime 2. The square marker and circular markers represent measurements for two similar prepared samples with different cantilevers. Triangular markers show a third set of measurements with focus on smaller islands using an alternative UHV-AFM set-up and improved sample preparation (see text for details). (b) Graph featuring 39 non-vanishing friction events under UHV conditions for particle sizes up to $310{,}000~\text{nm}^2$. The blue marker highlights the data point derived from Fig. 2(e). The data is well approximated by a linear fit with $F_f = (1.04 \pm 0.06)~\text{pN/nm}^2 \times A - (2.72 \pm 7.04)~\text{nN}$ (black solid lines) with no statistically significant offset. (c) Experimental data obtained under ambient conditions for particles with contact areas between $21{,}000~\text{nm}^2$ and $62{,}000~\text{nm}^2$. Two different regimes can once again be identified. The events featuring substantial friction follow an approximate linear dependency with $F_f = (40 \pm 1)~\text{pN/nm}^2 \times A - (48 \pm 74)~\text{nN}$ (black solid).

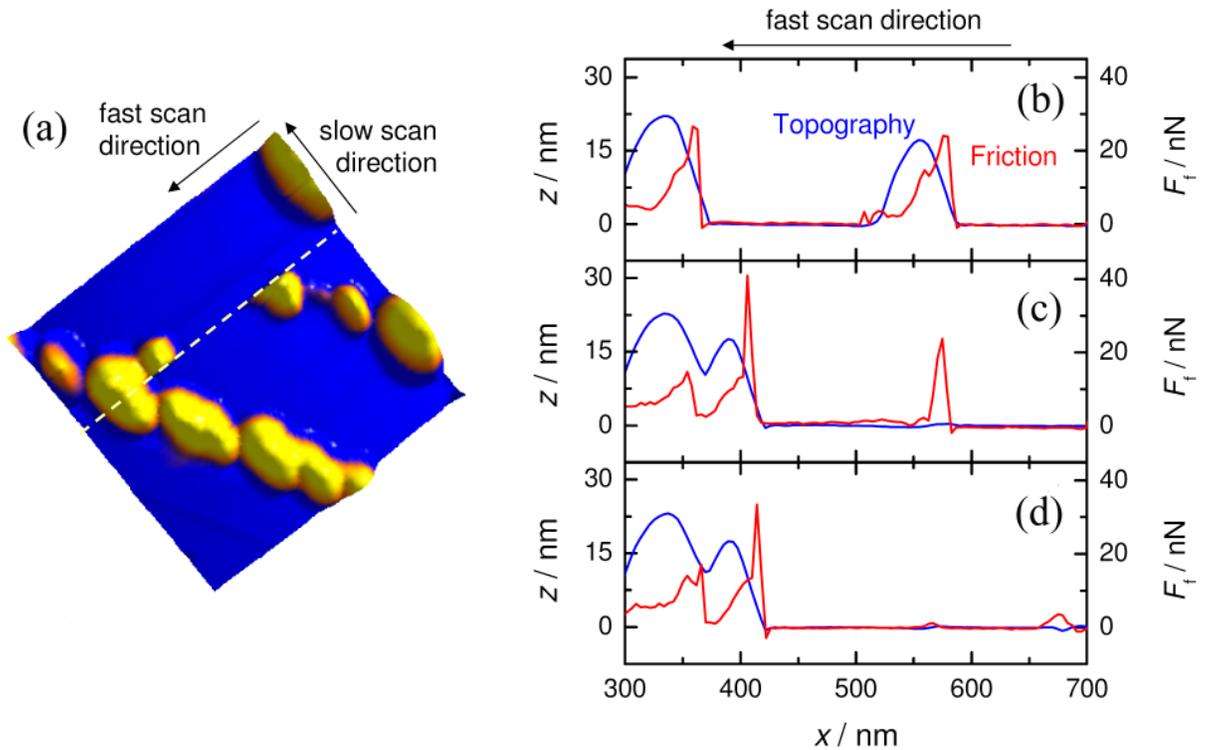

Figure 4: (a) Topographical scan during which a translation event of an Sb nanoparticle with a contact area of $8,000$ nm$^2$ took place. The particle was displaced during the recording of one single scan line (dashed line) and thus appears "cut". (b)-(d) The corresponding scan lines just before (b), during (c), and right after (d) the translation of the particle. In contrast to Fig. 2(e), the friction signal here only shows a peak where the tip hits the island at its initial position ($x = 580$ nm) and remains flat afterwards until the island reaches its new resting position at $x = 425$ nm. The frictional force during translation was well below $1$ nN.